# Dispersion of electromagnetic waves in linear, homogeneous, and isotropic media

M. Mansuripur
masud@optics.arizona.edu
James C. Wyant College of Optical Sciences, The University of Arizona, Tucson, Arizona 85721

**Abstract**. An electromagnetic wave-packet propagating in a linear, homogeneous, and isotropic medium changes shape while its envelope travels with different velocities at different points in spacetime. In general, a wave-packet can be described as a superposition of plane-waves having different frequencies $\omega$ and different propagation vectors ***k***. While the angular spread of the $k$-vectors gives rise to diffractive effects, it is the frequency-dependence of the refractive index of the host medium that is commonly associated with optical dispersion. When the spectral distribution of the wave-packet is confined to a narrow band of frequencies, and also when the spread of the $k$-vectors is not too broad, it is possible, under certain circumstances, to obtain analytical expressions for the local and/or global trajectory of the packet's envelope as it evolves in time. This paper is an attempt at a systematic description of the underlying physical assumptions and mathematical arguments leading to certain well-known properties of narrowband electromagnetic wave-packets in the presence of diffractive as well as (temporally) dispersive effects.

**1. Introduction**. It is well known that packets of electromagnetic (EM) waves propagating inside linear, homogenous, and isotropic (LHI) media exhibit dispersive as well as diffractive effects.[1-7] Such behavior is rooted in the underlying composition of a wave-packet as a superposition of EM plane-waves that occupy a band of frequencies, $\omega$, and possess a range of propagation vectors, ***k***, commonly known as $k$-vectors. The angular spread of the $k$-vectors is responsible for the diffractive effects, while the frequency-dependence of the optical properties of the host medium gives rise to dispersive phenomena such as wave-packet distortion and the emergence of the notion of group velocity as a distinct characteristic of the envelope of the wave-packet.[1,2,5]

Both the dielectric permittivity $\varepsilon$ and the magnetic permeability $\mu$ of the host medium can be functions of the oscillation frequency $\omega$ of the EM field. Thus, the refractive index $n$, which is readily shown to be equal to $\sqrt{\mu\varepsilon}$, is a function of $\omega$, and it is this parameter's frequency-dependence that is primarily responsible for the observed dispersive phenomena in optical systems.[1,2] We mention in passing that, in addition to their dependence on $\omega$, the $\mu$ and $\varepsilon$ of a material medium could also depend on the direction of the $k$-vector, thus producing an effect known as spatial dispersion.[8,9] In this paper, however, we shall stay away from the phenomena associated with spatial dispersion, focusing solely on temporal dispersive effects that arise from the dependence of the refractive index $n$ on the frequency $\omega$ of individual plane-waves.

In the next section, where the propagation of narrowband wave-packets within LHI media is examined, we derive a general formula for the trajectory of the packet envelope's peak traveling through spacetime. A special case of this analysis pertaining to wave-packets that enter a highly absorptive medium (e.g., a metallic mirror) is presented in Sec.3. Section 4 is devoted to an analysis of narrowband wave-packets traveling through transparent media subject to additional constraints that would arise when the packet is confined within a waveguide. A special case of confinement involving surface waves (e.g., surface plasmon polaritons[10-14]) is discussed in Sec.5. Finally, in Sec.6, we examine the problem of wave-packet propagation within an LHI medium from a somewhat different perspective, and arrive at a formula for the local group velocity of the packet's envelope at its various spatial locations.[1] Examples are provided throughout the paper to clarify some of the more abstract concepts and shed light on purely mathematical arguments. A summary of the main results and a few closing remarks form the subject of the final section.

**2. Wave-packet within a linear, homogeneous, and isotropic medium.** With reference to Fig.1, consider an LHI medium specified by its relative permittivity and permeability, $\varepsilon(\omega)$ and $\mu(\omega)$, both of which, in general, are complex functions of $\omega$. Limiting our attention to media that obey the Lorentz oscillator model of the refractive index[1-5] — and that do *not* exhibit spatial dispersion[8,9] — we learn from Maxwell's equations that $k^2 = (\omega/c)^2 \mu(\omega)\varepsilon(\omega)$. Let us further restrict our attention to semi-infinite media which have a flat interface with another medium, say, the $xy$-plane at $z = 0$, through which plane-waves with known values of $\omega$, $k_x$, and $k_y$ can enter our semi-infinite medium. Either a single plane-wave or many plane-waves may enter the medium, and we shall assume for each such plane-wave that the values of $\omega$, $k_x = (\omega/c)\sigma_x$, and $k_y = (\omega/c)\sigma_y$ are real but otherwise arbitrary. Each plane-wave will thus have a single $k_z = (\omega/c)\sqrt{\mu(\omega)\varepsilon(\omega) - \sigma_x^2 - \sigma_y^2}$, where the sign of the square root must subsequently be chosen to allow the plane-wave to propagate in the correct direction. Inside the medium, each plane-wave will have its electric field components $(E_x, E_y)$ determined by the boundary conditions at the $z = 0$ interface; its $E_z$ component, however, will be given by Maxwell's first equation, $\boldsymbol{k} \cdot \boldsymbol{E} = 0$, as $E_z = -(k_x E_x + k_y E_y)/k_z$. Also, the beam's magnetic field vector $\boldsymbol{H}$ will be fully determined by Maxwell's third equation, $\boldsymbol{H} = \boldsymbol{k} \times \boldsymbol{E}/\mu_0 \mu(\omega)\omega$.[2,5]

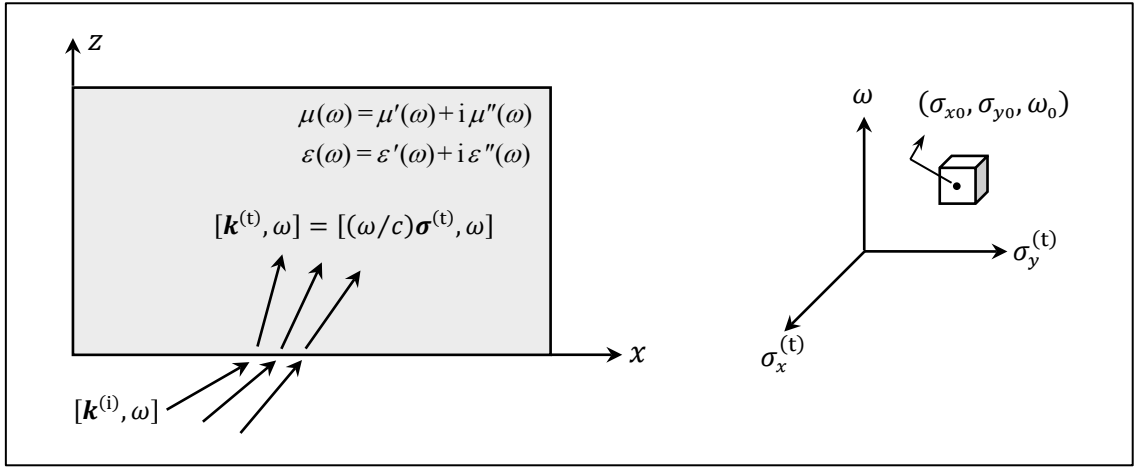

**Fig.1**. A wave-packet containing a narrow range of wave-vectors $\boldsymbol{k}$ and frequencies $\omega$ enters a semi-infinite medium. (The superscripts i and t refer to the incident and transmitted entities, respectively.) The LHI medium is specified by its relative permeability and permittivity, $\mu(\omega)$ and $\varepsilon(\omega)$. The small volume occupied by the transmitted beam within the $(\boldsymbol{k}, \omega)$ space is centered at $(\sigma_{x0}, \sigma_{y0}, \omega_0)$, as shown. Inside the medium, the plane-wave spectrum of the wave-packet is fully specified by its $E$-field amplitude distributions $\mathbb{E}_x(\sigma_x, \sigma_y, \omega)$ and $\mathbb{E}_y(\sigma_x, \sigma_y, \omega)$.

In general, $k_z$ will be complex and may, therefore, be written as $k_z' + \mathrm{i}k_z''$. Since $k_x, k_y$ are assumed to be real, the plane-wave's propagation direction will be $\boldsymbol{k} = k_x \hat{\boldsymbol{x}} + k_y \hat{\boldsymbol{y}} + k_z' \hat{\boldsymbol{z}}$, while its attenuation, with the decay coefficient $\exp(-k_z'' z)$, will take place along the $z$-axis.

Suppose that, instead of a single plane-wave, we have a wave-packet of finite duration as well as finite cross-sectional diameter in the $xy$-plane, yet the diameter is fairly large and the pulse duration is long, so that the field's spectral amplitude $\vec{\mathbb{E}}(\sigma_x, \sigma_y, \omega)$ in the $(\boldsymbol{k}, \omega)$ space is confined to a narrow range of values around $(\sigma_{x0}, \sigma_{y0}, \omega_0)$. We assume that $\mu(\omega) = \mu' + \mathrm{i}\mu''$ and $\varepsilon(\omega) = \varepsilon' + \mathrm{i}\varepsilon''$, with $0 \leq \mu'' \ll 1$ and also $0 \leq \varepsilon'' \ll 1$, so that the medium is weakly absorptive. The $z$-component of the $k$-vector may, therefore, be approximated as follows:

$$k_z = k_z' + ik_z'' \cong (\omega/c)(\mu'\varepsilon' - \mu''\varepsilon'' - \sigma_x^2 - \sigma_y^2)^{\frac{1}{2}} + i(\omega/c)\frac{\mu'\varepsilon'' + \varepsilon'\mu''}{2(\mu'\varepsilon' - \mu''\varepsilon'' - \sigma_x^2 - \sigma_y^2)^{\frac{1}{2}}}. \quad (1)$$

The implicit assumption here is that, not only is $\mu'\varepsilon'$ positive, but also it is greater than $\mu''\varepsilon'' + \sigma_x^2 + \sigma_y^2$, so that, for all $(\sigma_x, \sigma_y, \omega)$ values of interest in the following analysis, the square root appearing in Eq.(1) is real-valued. The sign of $k_z'$ is determined by the sign of the imaginary part of $\mu\varepsilon$, namely, $\mu'\varepsilon'' + \varepsilon'\mu''$. When both $\mu'$ and $\varepsilon'$ are positive, the medium will have a positive refractive index, in which case $k_z'$ and $k_z''$ will be positive. For a negative-index medium, however, both $\mu'$ and $\varepsilon'$ will be negative and, therefore, the sign of $k_z'$ must be reversed.[7]

Writing $\mathbb{E}_x(\sigma_x, \sigma_y, \omega) = |\mathbb{E}_x(\sigma_x, \sigma_y, \omega)| \exp[i\phi_x(\sigma_x, \sigma_y, \omega)]$, and likewise for $\mathbb{E}_y, \mathbb{E}_z$, we express each Cartesian component of the $E$-field as a superposition of all the various plane-waves that have entered the semi-infinite medium, as follows:

$$E_x(\mathbf{r},t) = \text{Re} \int_{\omega=0}^{\infty} \iint_{\sigma_x,\sigma_y} |\mathbb{E}_x(\sigma_x, \sigma_y, \omega)| \exp[i\phi_x(\sigma_x, \sigma_y, \omega)] \exp(-k_z'' z)$$
$$\times \exp[i(\omega/c)(\sigma_x x + \sigma_y y \pm \sqrt{\mu'\varepsilon' - \mu''\varepsilon'' - \sigma_x^2 - \sigma_y^2}\, z - ct)]\, d\sigma_x d\sigma_y d\omega. \quad (2)$$

In the above equation, the plus and minus signs appearing before the radical correspond, respectively, to positive- and negative-index media. The attenuation factor, $\exp(-k_z'' z)$, is likely to be fairly constant over the domain of integration (i.e., the small cube in Fig.1 that contains the significant values of $\sigma_x, \sigma_y, \omega$); as such, it may be evaluated at the mid-point $(\sigma_{x0}, \sigma_{y0}, \omega_0)$ of the integration range, then brought out of the integral.

Now, in the $(\mathbf{r}, t)$ space, wherever the phase of the remaining integrand in Eq.(2) happens to be more or less constant in the vicinity of $(\sigma_{x0}, \sigma_{y0}, \omega_0)$, one can say that the envelope of $E_x(\mathbf{r}, t)$ at that location is at or near the peak value for the entire wave-packet. The stationary point of the phase profile is determined by setting its partial derivatives with respect to $\sigma_x, \sigma_y, \omega$, evaluated at $(\sigma_{x0}, \sigma_{y0}, \omega_0)$, equal to zero; that is,

$$\frac{\partial \phi_x}{\partial \sigma_x} + \frac{\omega_0}{c}\left[x \mp \frac{z\sigma_{x0}}{(\mu'\varepsilon' - \mu''\varepsilon'' - \sigma_{x0}^2 - \sigma_{y0}^2)^{\frac{1}{2}}}\right] = 0, \quad (3a)$$

$$\frac{\partial \phi_x}{\partial \sigma_y} + \frac{\omega_0}{c}\left[y \mp \frac{z\sigma_{y0}}{(\mu'\varepsilon' - \mu''\varepsilon'' - \sigma_{x0}^2 - \sigma_{y0}^2)^{\frac{1}{2}}}\right] = 0, \quad (3b)$$

$$\frac{\partial \phi_x}{\partial \omega} + \frac{1}{c}(\sigma_{x0} x + \sigma_{y0} y \pm \sqrt{\mu'\varepsilon' - \mu''\varepsilon'' - \sigma_{x0}^2 - \sigma_{y0}^2}\, z - ct) \pm \left(\frac{z\omega_0}{c}\right)\frac{\partial(\mu'\varepsilon' - \mu''\varepsilon'')/\partial\omega}{2(\mu'\varepsilon' - \mu''\varepsilon'' - \sigma_{x0}^2 - \sigma_{y0}^2)^{\frac{1}{2}}} = 0. \quad (3c)$$

Solving these equations for $(x, y, z)$, we find the peak position of the pulse at time $t$, as follows:

$$x_{\text{peak}} = \frac{\pm \sigma_{x0}\, z_{\text{peak}}}{(\mu'\varepsilon' - \mu''\varepsilon'' - \sigma_{x0}^2 - \sigma_{y0}^2)^{\frac{1}{2}}} - \left(\frac{c}{\omega_0}\right)\frac{\partial \phi_x}{\partial \sigma_x}\bigg|_{\sigma_{x0},\sigma_{y0},\omega_0}, \quad (4a)$$

$$y_{\text{peak}} = \frac{\pm \sigma_{y0}\, z_{\text{peak}}}{(\mu'\varepsilon' - \mu''\varepsilon'' - \sigma_{x0}^2 - \sigma_{y0}^2)^{\frac{1}{2}}} - \left(\frac{c}{\omega_0}\right)\frac{\partial \phi_x}{\partial \sigma_y}\bigg|_{\sigma_{x0},\sigma_{y0},\omega_0}, \quad (4b)$$

$$z_{\text{peak}} = \frac{(\mu'\varepsilon' - \mu''\varepsilon'' - \sigma_{x0}^2 - \sigma_{y0}^2)^{\frac{1}{2}}}{\sqrt{\mu'\varepsilon' - \mu''\varepsilon''}} \times \frac{c[t - (\partial\phi_x/\partial\omega) + (\sigma_{x0}/\omega_0)(\partial\phi_x/\partial\sigma_x) + (\sigma_{y0}/\omega_0)(\partial\phi_x/\partial\sigma_y)]}{\pm(\sqrt{\mu'\varepsilon' - \mu''\varepsilon''} + \omega_0 \partial\sqrt{\mu'\varepsilon' - \mu''\varepsilon''}/\partial\omega)}\bigg|_{\sigma_{x0},\sigma_{y0},\omega_0}. \quad (4c)$$

The peak of the wave-packet, therefore, propagates along its path with the expected group velocity[1,2] $\pm c/(\sqrt{\mu'\varepsilon' - \mu''\varepsilon''} + \omega_0 \partial_\omega \sqrt{\mu'\varepsilon' - \mu''\varepsilon''})$, which must be positive for positive-index as well as negative-index media. As the packet propagates along its path through the medium, there will be attenuation, due to multiplication by the decay-factor $\exp(-k_z'' z)$, and distortion caused by dispersion as well as diffraction. Nonetheless, the peak position of the packet at time $t$ is generally expected to remain in the vicinity of the point specified by Eqs.(4). Similar relations exist for the $y$- and $z$-components of the $E$-field (and also for the $H$-field); their only distinction being the initial peak position of the packet for different field components, as determined, for example, in the case of $E_y$ and $E_z$, by the partial derivatives of $\phi_y(\sigma_x, \sigma_y, \omega)$ and $\phi_z(\sigma_x, \sigma_y, \omega)$.

**Example**. A chirped Gaussian light pulse of width $T$, center frequency $\omega_0$, and quadratic chirp coefficient $\alpha_0$, is linearly polarized along the $x$-axis. In its cross-sectional $xy$-plane, the packet diameter is large and its wavefront is fairly uniform, so the angular spread of the $k$-vectors can be ignored. In the $xy$-plane at $z = 0$, the $E$-field of the pulse has the following time-dependence:

$$\boldsymbol{E}_0(t) = E_{x0} e^{-(t/T)^2} \cos(\omega_0 t + \alpha_0 t^2)\, \widehat{\boldsymbol{x}}. \tag{5}$$

Considering that the Fourier transform of $e^{-(at^2 + bt)}$ is $(\pi/a)^{1/2} e^{-(\omega + ib)^2/4a}$, one may write the $E$-field distribution of Eq.(5) as the real part of an inverse Fourier transform, as follows:

$$\boldsymbol{E}_0(t) = \mathrm{Re}\left[\frac{T}{\sqrt{4\pi(1 + i\alpha_0 T^2)}} \int_{-\infty}^{\infty} e^{-T^2(\omega - \omega_0)^2/[4(1 + i\alpha_0 T^2)]} e^{-i\omega t} d\omega\right] E_{x0} \widehat{\boldsymbol{x}}. \tag{6}$$

For sufficiently large $\omega_0$, the contribution of negative frequencies to the above integral can be ignored, thus allowing the lower limit of the integral to be raised from $-\infty$ to zero. Since the wave-packet is assumed to be propagating along the positive $z$-axis, the $E$-field in the $xy$-plane at an arbitrary $z$ may now be obtained by multiplying $e^{ikz}$ into the integrand of Eq.(6); that is,

$$\boldsymbol{E}(z, t) \cong \mathrm{Re}\left[\frac{T}{\sqrt{4\pi(1 + i\alpha_0 T^2)}} \int_{\omega=0}^{\infty} e^{-T^2(\omega - \omega_0)^2/[4(1 + i\alpha_0 T^2)]} e^{i(kz - \omega t)} d\omega\right] E_{x0} \widehat{\boldsymbol{x}}. \tag{7}$$

In a transparent medium of refractive index $n(\omega)$, the magnitude of the $k$-vector is given by $k = \omega n(\omega)/c$. Assuming a narrowband wave-packet, we proceed to approximate $kz - \omega t$ using a Taylor series expansion up to the 2nd order around the center frequency $\omega_0$. We will have

$$kz - \omega t = \omega n(\omega) z/c - \omega t \cong [\omega_0 n(\omega_0) z/c] + \partial_\omega [\omega n(\omega)/c]_{\omega_0} (\omega - \omega_0) z$$
$$+ \partial_\omega^2 [\omega n(\omega)/2c]_{\omega_0} (\omega - \omega_0)^2 z - [\omega_0 t + (\omega - \omega_0) t]. \tag{8}$$

Our notation will be simplified if we introduce the group velocity $v_g = c/\partial_\omega [\omega n(\omega)]_{\omega_0}$ and the auxiliary coefficient $\beta = 2\partial_\omega^2 [\omega n(\omega)/c]_{\omega_0}$. Substituting Eq.(8) into Eq.(7) and carrying out the integral now yields

$$\boldsymbol{E}(z, t) \cong \mathrm{Re}\left\{\frac{\exp\{i(\omega_0/c)[n(\omega_0) z - ct]\}}{\sqrt{1 + (\alpha_0 - iT^{-2})\beta z}} \exp\left[-\frac{(1 + i\alpha_0 T^2)[(z/v_g) - t]^2}{T^2 + (\alpha_0 T^2 - i)\beta z}\right]\right\} E_{x0} \widehat{\boldsymbol{x}}. \tag{9}$$

Rearranging and streamlining the above equation, we finally arrive at

$$\boldsymbol{E}(z, t) \cong \mathrm{Re}\left\{\sqrt{T/T(z)}\, e^{i\varphi(z)} e^{-[T^{-2}(z) + i\alpha(z)][(z/v_g) - t]^2} e^{-i\omega_0 t}\right\} E_{x0} \widehat{\boldsymbol{x}}, \tag{10}$$

where,

$$\varphi(z) = \omega_0 n(\omega_0)(z/c) + \tan^{-1}\{\beta z / [2T^2(1 + \alpha_0 \beta z)]\}, \tag{10a}$$

$$T(z) = T\sqrt{(1 + \alpha_0 \beta z)^2 + (\beta z/T^2)^2}, \tag{10b}$$

$$\alpha(z) = \frac{(1+\alpha_0 \beta z)\alpha_0 + (\beta z/T^4)}{(1+\alpha_0 \beta z)^2 + (\beta z/T^2)^2}. \tag{10c}$$

It is seen that, at any given distance $z$ from the origin, the Gaussian character of the pulse (as a function of time $t$) is preserved.[2,6] While the peak of the Gaussian propagates with the group velocity $v_g \hat{z}$, the pulse width $T$ and the quadratic chirp coefficient $\alpha$ vary as functions of $z$. (Jackson[2] presents a parallel analysis involving a wave-packet that is a Gaussian function of $z$.)

**3. Strongly attenuating medium**. Returning to Eq.(2), let us consider the case of a strongly attenuating medium, where $k_z'' \gg 0$. This could happen, for instance, when $\mu''$ and/or $\varepsilon''$ are large, or when either $\mu'$ or $\varepsilon'$ (but not both) is large and negative. Under such circumstances, the penetration depth of the beam into the LHI medium is rather shallow, and one may analyze the behavior of the wave-packet immediately beneath the entrance facet at $z = 0^+$. We will have

$$E_x(x, y, 0^+, t) = \text{Re} \int_{\omega=0}^{\infty} \iint_{\sigma_x, \sigma_y} |\mathbb{E}_x(\sigma_x, \sigma_y, \omega)| e^{i[\phi_x(\sigma_x, \sigma_y, \omega) + (\omega/c)(\sigma_x x + \sigma_y y - ct)]} d\sigma_x d\sigma_y d\omega. \tag{11}$$

The peak of the packet's envelope arrives beneath the surface of the LHI medium when the phase of the integrand in Eq.(11) is more or less constant. This we determine by setting the partial derivatives of the phase with respect to $(\sigma_x, \sigma_y, \omega)$ equal to zero. We find

$$x_{\text{peak}} = -\left(\frac{c}{\omega_0}\right) \frac{\partial \phi_x}{\partial \sigma_x}\bigg|_{\sigma_{x0}, \sigma_{y0}, \omega_0}, \tag{12a}$$

$$y_{\text{peak}} = -\left(\frac{c}{\omega_0}\right) \frac{\partial \phi_x}{\partial \sigma_y}\bigg|_{\sigma_{x0}, \sigma_{y0}, \omega_0}, \tag{12b}$$

$$t_{\text{peak}} = \frac{\partial \phi_x}{\partial \omega} - \left(\frac{\sigma_{x0}}{\omega_0}\right) \frac{\partial \phi_x}{\partial \sigma_x} - \left(\frac{\sigma_{y0}}{\omega_0}\right) \frac{\partial \phi_x}{\partial \sigma_y}\bigg|_{\sigma_{x0}, \sigma_{y0}, \omega_0}. \tag{12c}$$

This is all that one can say about the peak position of the packet just beneath the surface, namely its location and its time of arrival. Of course, at any given time while the pulse lasts, there will be one or more local maxima of the field at or beneath the surface, but these maxima are not associated with a stationary phase of the integrand in Eq.(11).

**Example**. Let the incident wave-packet in the system depicted in Fig.1 be specified as

$$E_y^{(i)}(x, z = 0^-, t) = \text{Re} \int_{\omega=0}^{\infty} \int_{\sigma_x} \mathbb{E}_y^{(i)}(\sigma_x, \omega) e^{i(\omega/c)(\sigma_x x - ct)} d\sigma_x d\omega. \tag{13}$$

This $s$-polarized packet has a finite width along the $x$-axis and a finite duration, with a (generally complex) spatio-temporal spectrum $\mathbb{E}_y^{(i)}(\sigma_x, \omega) = |\mathbb{E}_y^{(i)}| e^{i\phi_y}$. Suppose the transparent medium of incidence has the real and positive refractive index $n(\omega)$, while the metallic mirror into which the light is partially transmitted has a large, negative, real-valued permittivity $\varepsilon_0 \varepsilon(\omega)$. The Fresnel transmission coefficient at the $z = 0$ interface is known to be[1,2,7]

$$\tau(\sigma_x, \omega) = \mathbb{E}_y^{(t)}/\mathbb{E}_y^{(i)} = \frac{2[n^2(\omega) - \sigma_x^2]^{1/2}}{[n^2(\omega) - \sigma_x^2]^{1/2} + i[\sigma_x^2 - \varepsilon(\omega)]^{1/2}}. \tag{14}$$

Thus, the transmitted $E$-field immediately beneath the mirror's surface is given by

$$E_y^{(t)}(x,y,z=0^+,t) = \mathrm{Re}\int_{\omega=0}^{\infty}\int_{\sigma_x}|\tau(\sigma_x,\omega)\mathbb{E}_y^{(i)}(\sigma_x,\omega)|$$
$$\times \exp\mathrm{i}\{\phi_y(\sigma_x,\omega) - \tan^{-1}\sqrt{[n^2(\omega)-\sigma_x^2]/[\sigma_x^2-\varepsilon(\omega)]} + (\omega/c)(\sigma_x x - ct)\}\mathrm{d}\sigma_x\mathrm{d}\omega. \quad (15)$$

The phase of the integrand in the above equation may now be used to determine the peak position $(x_{\mathrm{peak}}, t_{\mathrm{peak}})$ of the $E$-field's envelope within the skin-depth of the metallic mirror.

**4. Propagation inside a waveguide**. An interesting situation arises when the host medium is somehow "selective," in the sense that, for each frequency $\omega$, there exist only a few specific values of $k_x$ and/or $k_y$ that can enter (or survive within) the medium. The waveguide depicted in Fig.2 is an example of such selective media. Let us assume that, in a particular situation, the guided mode consists only of the following four $k$-vectors for each value of $\omega$:

$$\boldsymbol{k}(\omega) = [\pm k_x(\omega), \pm k_y(\omega), k_z(\omega) = \sqrt{(\omega/c)^2\mu(\omega)\varepsilon(\omega) - k_x^2 - k_y^2}]. \quad (16\mathrm{a})$$

To simplify the argument, we assume that $k_x, k_y$ are real-valued. Furthermore, the complex-amplitude, say, $\mathbb{E}_x$, of each component of the $E$-field associated with the above $k$-vectors is assumed to be given by

$$\mathbb{E}_x(\pm k_x, \pm k_y, \omega) = |\mathbb{E}_x(\omega)|\exp\{\mathrm{i}[\phi_x(\omega) \pm \phi_{xx}(\omega) \pm \phi_{xy}(\omega)]\}. \quad (16\mathrm{b})$$

For the weakly-absorbing medium with a positive refractive index that is presently under consideration, the guided wave-packet's $E_x$-field may thus be written as

$$E_x(\boldsymbol{r},t) = \mathrm{Re}\int_0^{\infty} 4|\mathbb{E}_x(\omega)|\exp[\mathrm{i}\phi_x(\omega)]\cos[k_x(\omega)x + \phi_{xx}(\omega)]\cos[k_y(\omega)y + \phi_{xy}(\omega)]$$
$$\times \exp[-k_z''(\omega)z]\exp\{\mathrm{i}[\sqrt{(\omega/c)^2(\mu'\varepsilon' - \mu''\varepsilon'') - k_x^2 - k_y^2}\,z - \omega t]\}\mathrm{d}\omega. \quad (17)$$

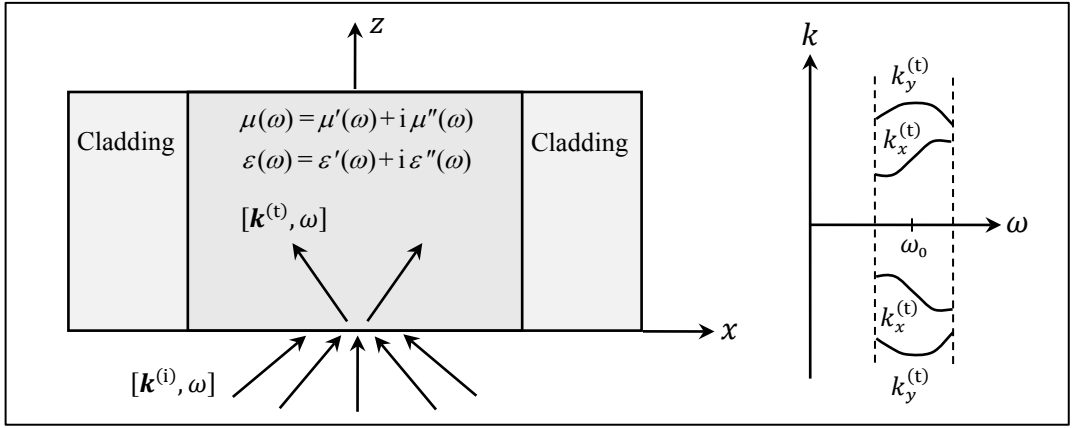

**Fig.2**. A waveguide is selective in its acceptance of the incident wave-vectors. It also could modify the accepted $k$-vectors (or introduce new ones) as a result of interactions between the admitted wave-packet and the cladding(s). The guiding medium in the present example is assumed to have a positive refractive index and weak absorption, that is, $0 \le \mu'' \ll \mu'$ and $0 \le \varepsilon'' \ll \varepsilon'$. Typically, for each frequency $\omega$ that is present in the incident packet, certain well-defined values of $k_x$ and $k_y$ propagate in the guiding region. The inset shows a specific example in which, for each incident frequency $\omega$, four wave-vectors are admitted into the guiding medium; the admitted wave-vectors are $\boldsymbol{k}(\omega) = [\pm k_x(\omega), \pm k_y(\omega), k_z(\omega) = \sqrt{(\omega/c)^2\mu(\omega)\varepsilon(\omega) - k_x^2 - k_y^2}]$. To simplify the argument, $k_x$ and $k_y$ are assumed to be real. The actual dependence of $k_x$ and $k_y$ on $\omega$ is determined by the geometry of the waveguide as well as the EM properties of the core and cladding materials.

Here, $k_z''(\omega) \cong \tfrac{1}{2}(\omega/c)^2(\mu'\varepsilon'' + \varepsilon'\mu'')/\sqrt{(\omega/c)^2(\mu'\varepsilon' - \mu''\varepsilon'') - k_x^2 - k_y^2}$ is the imaginary part of the propagating wave's $k$-vector, obtained in the limit of weak absorption.

To determine the peak position of the wave-packet as a function of time, we note that the phase of the narrowband integrand in Eq.(17) will be essentially constant if its derivative with respect to $\omega$ vanishes. This yields,

$$z_{\text{peak}} = \left.\frac{t - (\partial \phi_x / \partial \omega)}{\partial_\omega [(\omega/c)^2(\mu'\varepsilon' - \mu''\varepsilon'') - k_x^2 - k_y^2]^{\tfrac{1}{2}}}\right|_{\omega=\omega_0}. \qquad (18)$$

Thus, the group velocity of the guided packet is seen to be not only a function of $\varepsilon(\omega)$ and $\mu(\omega)$ of the waveguide's core material, but it also depends on the frequency-dependence of the values of $k_x$ and $k_y$ that are admitted into the waveguide.

**Example**. Consider a perfectly electrically conducting hollow cylinder of radius $R$, filled with a homogeneous and isotropic dielectric material of permeability $\mu_0\mu(\omega)$ and permittivity $\varepsilon_0\varepsilon(\omega)$. The dielectric host medium being transparent, its refractive index $n(\omega) = \sqrt{\mu(\omega)\varepsilon(\omega)}$ is a positive real-valued function of the frequency $\omega$.

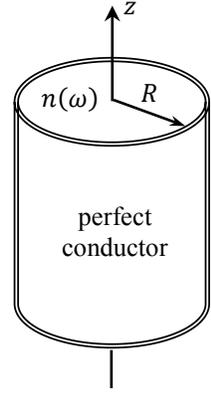

In this cylindrically symmetric system, the $k$-vector is written as $\boldsymbol{k} = k_0(\sigma_r\hat{\boldsymbol{r}} + \sigma_z\hat{\boldsymbol{z}})$, where $k_0 = \omega/c$ is the wave-number in free space, and $\sigma_r^2 + \sigma_z^2 = n^2(\omega)$. Recalling that the speed of light in vacuum is $c = (\mu_0\varepsilon_0)^{-\tfrac{1}{2}}$ and that the impedance of free space is $Z_0 = (\mu_0/\varepsilon_0)^{\tfrac{1}{2}}$, one can show that the solution of Maxwell's equations for the transverse magnetic (TM) guided modes of the EM field inside the cylinder is given by[2,7,15]

$$\boldsymbol{E}(\boldsymbol{r},t) = [(\mathrm{i}\sigma_z/\sigma_r)J_m'(k_0\sigma_r r)\hat{\boldsymbol{r}} - (m\sigma_z/k_0 r\sigma_r^2)J_m(k_0\sigma_r r)\hat{\boldsymbol{\varphi}} + J_m(k_0\sigma_r r)\hat{\boldsymbol{z}}]e^{\mathrm{i}m\varphi}e^{\mathrm{i}(k_0\sigma_z z-\omega t)}. \quad (19\text{a})$$

$$Z_0\boldsymbol{H}(\boldsymbol{r},t) = [(m\varepsilon/k_0 r\sigma_r^2)J_m(k_0\sigma_r r)\hat{\boldsymbol{r}} + (\mathrm{i}\varepsilon/\sigma_r)J_m'(k_0\sigma_r r)\hat{\boldsymbol{\varphi}}]e^{\mathrm{i}m\varphi}e^{\mathrm{i}(k_0\sigma_z z-\omega t)}. \quad (19\text{b})$$

Here, the integer $m$ is the azimuthal mode number, $J_m(\cdot)$ is a Bessel function of the first kind, order $m$, and $J_m'(\cdot)$ is the derivative of $J_m(\cdot)$ with respect to its argument. The boundary conditions at $r = R$, namely $E_z = E_\varphi = H_r = 0$, are imposed by the fact that, at the interior facet of the perfect conductor, the tangential components of the $E$-field as well as the perpendicular component of the $B$-field must vanish. This requires that $k_0\sigma_r R$ be a zero of $J_m(\cdot)$. Denoting by $\rho_{mv}$ the $v^{\text{th}}$ zero of $J_m(\rho)$, we find $\sigma_r(\omega) = c\rho_{mv}/R\omega$. For a single-mode wave-packet having a narrow bandwidth $\Delta\omega$ centered at $\omega = \omega_0$, the stationary phase condition is given by

$$\partial(k_0\sigma_z z - \omega t)/\partial\omega = 0 \quad \rightarrow \quad \partial_\omega[(\omega/c)\sqrt{n^2(\omega) - \sigma_r^2(\omega)}\, z - \omega t]_{\omega=\omega_0} = 0. \qquad (20)$$

Equation (20) is readily solved to yield the group velocity of the wave-packet, as follows:

$$v_g = \frac{c\sqrt{1 - [\rho_{mv}\lambda_0/2\pi R n(\omega_0)]^2}}{n(\omega_0) + \omega_0 n'(\omega_0)}. \qquad (21)$$

Here, $\lambda_0 = 2\pi c/\omega_0$ is the vacuum wavelength of the guided mode at its center frequency, and the direction of $\boldsymbol{v}_g$ is along the $z$-axis.

**5. Dispersive propagation of surface waves**. Let us consider a surface wave-packet that has entered a "selective" medium in which each $\omega$ is allowed only a single $k$-vector; that is,

$$\boldsymbol{k}(\omega) = \left[k_x(\omega), k_y(\omega), k_z(\omega) = \sqrt{(\omega/c)^2\mu(\omega)\varepsilon(\omega) - k_x^2 - k_y^2}\right]. \qquad (22)$$

Moreover, we assume that the imaginary part $k_z''(\omega)$ of $k_z(\omega)$ is relatively large, so that the wave-packet is more or less confined to the surface region in the vicinity of $z = 0^+$. This could happen, for instance, when $\mu''$ and/or $\varepsilon''$ are large, or when either $\mu'$ or $\varepsilon'$ (but not both) is negative. The surface profile of any field component, say, $E_x$, may thus be written

$$E_x(x, y, z = 0^+, t) = \mathrm{Re} \int_0^\infty |\mathbb{E}_x(\omega)| \exp\{\mathrm{i}[\phi_x(\omega) + k_x(\omega)x + k_y(\omega)y - \omega t]\} \, \mathrm{d}\omega. \qquad (23)$$

The peak of the wave-packet occurs where the phase of the integrand in Eq.(23) happens to be stationary. To determine the peak position, we evaluate the derivative of the phase with respect to $\omega$ at $\omega = \omega_0$, then set it equal to zero, to find

$$(\partial k_x/\partial \omega) x_{\mathrm{peak}} + (\partial k_y/\partial \omega) y_{\mathrm{peak}} = t - (\partial \phi_x/\partial \omega)|_{\omega = \omega_0}. \qquad (24)$$

This shows the peak of the packet as a straight line in the $xy$-plane, always perpendicular to the vector $(\partial k_x/\partial \omega)|_{\omega_0} \hat{\boldsymbol{x}} + (\partial k_y/\partial \omega)|_{\omega_0} \hat{\boldsymbol{y}}$, while moving along the direction of this vector as time progresses. According to Eq.(23), there will be no loss of energy over time, which would be fine if the medium happens to be lossless, but is otherwise unacceptable. Loss may be introduced into the wave-packet by allowing $k_x$ and/or $k_y$ to have imaginary components.

Another concern with the wave-packet of Eq.(23) is that it is not laterally confined (i.e., in the direction perpendicular to its axis of propagation). Confinement can be produced by bringing in additional wave-vectors; for instance, by allowing each frequency $\omega$ to have two admissible $k$-vectors, as follows:

$$\boldsymbol{k}(\omega) = [k_x(\omega), \pm k_y(\omega), k_z(\omega) = \sqrt{(\omega/c)^2 \mu(\omega)\varepsilon(\omega) - k_x^2 - k_y^2}]. \qquad (25)$$

Furthermore, let the corresponding field amplitudes be $|\mathbb{E}_x(\omega)| \exp\{\mathrm{i}[\phi_x(\omega) \pm \phi_{xy}(\omega)]\}$. The field profile at the $z = 0^+$ surface will then become

$$E_x(x, y, z = 0^+, t) = \mathrm{Re} \int_0^\infty 2|\mathbb{E}_x(\omega)| \cos[k_y(\omega)y + \phi_{xy}(\omega)] \exp\{\mathrm{i}[\phi_x(\omega) + k_x(\omega)x - \omega t]\} \, \mathrm{d}\omega. \qquad (26)$$

The peak of this wave-packet will now move along the $x$-axis in accordance with Eq.(24), provided of course that $y_{\mathrm{peak}}$ is set equal to zero.

**Example**. Surface plasmon polaritons[10-14] are EM waves that reside at the interface between a metal and a dielectric, as shown in Fig.3. We examine the ideal case of a flat interface between two semi-infinite media, one being a transparent dielectric (glass) of refractive index $n(\omega)$, which is a real and positive entity, the other a lossless metal whose relative permittivity $\varepsilon(\omega)$ is real and negative. At optical frequencies, both material media can be reasonably expected to have a relative permeability of $\mu(\omega) = 1$. Assuming $k_x$ is real and positive, and $k_y = 0$, the dispersion relation $k^2 = (\omega/c)^2 \mu(\omega)\varepsilon(\omega)$ yields

$$k_z^{\mathrm{glass}} = -\mathrm{i}\sqrt{k_x^2 - (\omega/c)^2 n^2(\omega)}, \qquad k_z^{\mathrm{metal}} = \mathrm{i}\sqrt{k_x^2 - (\omega/c)^2 \varepsilon(\omega)}. \qquad (27)$$

The $p$-polarized waves in the two media have $E$-field components along both $x$- and $z$-axes, and magnetic field components along the $y$-axis, with $E_x$ and $H_y$ being continuous at the interface.[1,2,7] Maxwell's first equation, $\boldsymbol{k} \cdot \boldsymbol{E} = 0$, relates $E_z$ to $E_x$ via $E_z = -k_x E_x/k_z$, while his third equation, $\boldsymbol{k} \times \boldsymbol{E} = \mu_0 \mu(\omega)\omega \boldsymbol{H}$, yields $H_y = (k_z E_x - k_x E_z)/\mu_0 \omega$. We thus have

$$\boldsymbol{E}_{\mathrm{glass}}(\boldsymbol{r}, t) = E_{x0}[\hat{\boldsymbol{x}} - (k_x/k_z^{\mathrm{glass}})\hat{\boldsymbol{z}}] \exp[\mathrm{i}(k_x x + k_z^{\mathrm{glass}} z - \omega t)]. \qquad (28\mathrm{a})$$

$$\boldsymbol{H}_{\text{glass}}(\boldsymbol{r},t) = [\varepsilon_0 n^2(\omega)\,\omega E_{x0}/k_z^{\text{glass}}]\widehat{\boldsymbol{y}}\exp[i(k_x x + k_z^{\text{glass}} z - \omega t)]. \qquad (28\text{b})$$

$$\boldsymbol{E}_{\text{metal}}(\boldsymbol{r},t) = E_{x0}[\widehat{\boldsymbol{x}} - (k_x/k_z^{\text{metal}})\widehat{\boldsymbol{z}}]\exp[i(k_x x + k_z^{\text{metal}} z - \omega t)]. \qquad (29\text{a})$$

$$\boldsymbol{H}_{\text{metal}}(\boldsymbol{r},t) = [\varepsilon_0 \varepsilon(\omega)\,\omega E_{x0}/k_z^{\text{metal}}]\widehat{\boldsymbol{y}}\exp[i(k_x x + k_z^{\text{metal}} z - \omega t)]. \qquad (29\text{b})$$

Note that our choice of the $E$-field amplitudes ensures the continuity of $E_x$ at the interface, where $z = 0$. The remaining boundary condition (i.e., the continuity of $H_y$), forces $k_x$ to satisfy the following equation:

$$k_x = \pm \frac{(\omega/c)n(\omega)}{\sqrt{1 + n^2(\omega)/\varepsilon(\omega)}}. \qquad (30)$$

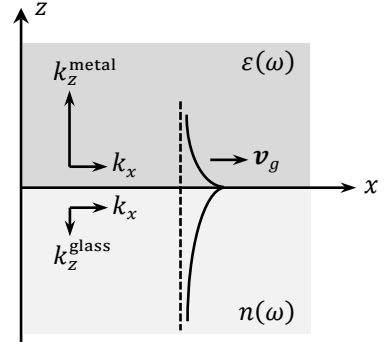

**Fig. 3**. A surface plasmon polariton resides at the interface between a transparent dielectric medium (glass) of refractive index $n(\omega)$ and a lossless metallic medium of permittivity $\varepsilon_0\varepsilon(\omega)$. At optical frequencies, the relative permeability of the media is expected to be $\mu(\omega) \cong 1$. Both $n(\omega)$ and $\varepsilon(\omega)$ are real-valued, with $n(\omega)$ being positive while $\varepsilon(\omega)$ is large and negative. Maxwell's boundary conditions require that $k_x$, $E_{x0}$, and $H_{y0}$ be the same in the two media. Both $k_z^{\text{glass}}$ and $k_z^{\text{metal}}$ are purely imaginary. A wave-packet formed by a superposition of monochromatic surface plasmons, having center frequency $\omega_0$ and a narrow bandwidth $\Delta\omega$, propagates along the $x$-axis with the group velocity $\boldsymbol{v}_g = \widehat{\boldsymbol{x}}/(\partial k_x/\partial \omega)|_{\omega_0}$.

Considering that $n(\omega)$ is real and positive, while $\varepsilon(\omega)$ is large, real, and negative, it is seen that $k_x$ will be real, with a magnitude greater than $\omega n(\omega)/c$, which renders the inhomogeneous plane-wave inside the glass medium an evanescent wave. Since both $k_z^{\text{glass}}$ and $k_z^{\text{metal}}$ are purely imaginary, the phase of a narrowband wave-packet (center frequency $= \omega_0$, bandwidth $= \Delta\omega$) will be given by $k_x(\omega) - \omega t$, which yields the group velocity $\boldsymbol{v}_g = \widehat{\boldsymbol{x}}/(\partial k_x/\partial \omega)|_{\omega=\omega_0}$.

**6. Local group velocity of wave-packets**. Consider a narrowband EM wave-packet whose scalar amplitude profile in the $xy$-plane at $z = 0$ is given by the real-valued function $a_0(x, y, t)$. Denoting the Fourier transform of this initial profile by $\mathbb{A}(k_x, k_y, \omega)$, we will have

$$a(\boldsymbol{r},t) = \text{Re}\Big\{(2\pi)^{-3}\int_{\omega=0}^{\infty}\iint_{k_x,k_y=-\infty}^{\infty}\mathbb{A}(k_x,k_y,\omega)\exp[i(k_x x + k_y y + k_z z - \omega t)]\,\mathrm{d}k_x\mathrm{d}k_y\mathrm{d}\omega\Big\}. \qquad (31)$$

Needless to say, the Fourier spectrum of the real-valued $a_0(x, y, t)$ must satisfy

$$\mathbb{A}(-k_x, -k_y, -\omega) = \mathbb{A}^*(k_x, k_y, \omega). \qquad (32)$$

Also, assuming the wave-packet propagates within a transparent LHI medium of refractive index $n(\omega)$, the dispersion relation yields

$$k_z = (\omega/c)\sqrt{n^2(\omega) - (ck_x/\omega)^2 - (ck_y/\omega)^2}. \qquad (33)$$

The following analysis will be streamlined if we substitute the complex function $A(\boldsymbol{r}, \omega)$ for the inner integral over the $k_x k_y$-plane that appears in Eq.(31), namely,

$$A(\boldsymbol{r},\omega) = |A(\boldsymbol{r},\omega)|e^{i\varphi(\boldsymbol{r},\omega)} = (2\pi)^{-3}\iint_{k_x,k_y}\mathbb{A}(k_x,k_y,\omega)e^{i\boldsymbol{k}\cdot\boldsymbol{r}}\mathrm{d}k_x\mathrm{d}k_y. \qquad (34)$$

We thus have

$$a(\pmb{r},t) = \text{Re} \int_{\omega=0}^{\infty} A(\pmb{r},\omega)e^{-i\omega t}d\omega = \text{Re} \int_{\omega=0}^{\infty} |A(\pmb{r},\omega)|e^{i[\varphi(\pmb{r},\omega)-\omega t]}d\omega. \tag{35}$$

Further below, we shall define a carrier frequency and phase $(\omega_c, \varphi_c)$ for the above wave-packet, but, quite independently of our choice of $(\omega_c, \varphi_c)$, the packet's envelope is given by

$$\mathcal{E}(\pmb{r},t) = \left| \int_{\omega=0}^{\infty} A(\pmb{r},\omega)e^{-i\omega t}d\omega \right|. \tag{36}$$

Given this real-valued and non-negative function of the spacetime coordinates, the squared magnitude of the envelope is readily expressed as follows:

$$\mathcal{E}^2(\pmb{r},t) = \int_{\omega=0}^{\infty} \int_{\omega'=0}^{\infty} A(\pmb{r},\omega)A^*(\pmb{r},\omega')e^{-i(\omega-\omega')t}d\omega d\omega'. \tag{37}$$

To specify the carrier frequency and phase, we return to Eq.(35) and identify a center frequency for the narrowband function $A(\pmb{r},\omega)$. Ideally, $A(\pmb{r},\omega)$ would have the form of a Hermitian function of $\omega$ (relative to the mid-point $\omega_c$ of the bandwidth) multiplied by a constant phase-factor $e^{i\varphi_c}$. This would make $(\omega_c, \varphi_c)$ the ideal frequency and phase of the carrier — i.e., render the carrier free of any residual phase modulation. In general, however, one cannot expect to have such an ideal $A(\pmb{r},\omega)$, and the best choice for $(\omega_c, \varphi_c)$ at any given point $\pmb{r}$ is one that would make $A(\pmb{r}, \omega - \omega_c)\exp(-i\varphi_c)$ as close to a Hermitian function of $\omega$ as possible. If it now happens, to a good approximation at least, that the phase $\varphi(\pmb{r},\omega)$ of $A(\pmb{r},\omega)$ is a linear function of $\omega$ within the bandwidth $\Delta\omega$ of the wave-packet, we may write

$$\varphi(\pmb{r},\omega) \cong \varphi(\pmb{r},\omega_c) + \partial_\omega \varphi(\pmb{r},\omega)|_{\omega=\omega_c}(\omega - \omega_c). \tag{38}$$

Here, $\varphi(\pmb{r},\omega_c) = \varphi_c$, while $\partial_\omega \varphi(\pmb{r},\omega_c)$ is the slope of the local phase profile of the wave-packet at the given point $\pmb{r}$. The approximate form of the wave-packet of Eq.(35), with its carrier explicitly separated from the envelope, may now be written as

$$a(\pmb{r},t) \cong \text{Re}\left\{ e^{i(\varphi_c - \omega_c t)} \int_{\Delta\omega} |A(\pmb{r},\omega)|e^{i(\omega-\omega_c)[\partial_\omega \varphi(\pmb{r},\omega_c)-t]}d\omega \right\}. \tag{39}$$

Similarly, substitution from Eq.(38) into Eq.(37) yields

$$\mathcal{E}^2(\pmb{r},t) \cong \iint_{\Delta\omega,\Delta\omega'} |A(\pmb{r},\omega)A(\pmb{r},\omega')|e^{i(\omega-\omega')[\partial_\omega \varphi(\pmb{r},\omega_c)-t]}d\omega d\omega'$$

$$= \iint_{\Delta\omega,\Delta\omega'} |A(\pmb{r},\omega)A(\pmb{r},\omega')| \cos\{(\omega-\omega')[\partial_\omega \varphi(\pmb{r},\omega_c) - t]\} d\omega d\omega'. \tag{40}$$

In the above equation, the knowledge that $\mathcal{E}^2(\pmb{r},t)$ is real has enabled us to replace the integrand's complex exponential factor with its real part.

Now, in going from $\pmb{r}$ to a nearby point $\pmb{r} + \Delta\pmb{r}$, the change in $\partial_\omega \varphi(\pmb{r},\omega_c)$ will be $\pmb{\nabla}[\partial_\omega \varphi(\pmb{r},\omega_c)] \cdot \Delta\pmb{r}$. Assuming $A(\pmb{r},\omega)$ varies slowly with $\pmb{r}$, setting $\pmb{\nabla}[\partial_\omega \varphi(\pmb{r},\omega_c)] \cdot \Delta\pmb{r} = \Delta t$ ensures that $\mathcal{E}(\pmb{r} + \Delta\pmb{r}, t + \Delta t) \cong \mathcal{E}(\pmb{r},t)$. For a fixed $\Delta t$, the shortest possible distance for the envelope to propagate away from $\pmb{r}$ would be along the direction of $\pmb{\nabla}[\partial_\omega \varphi(\pmb{r},\omega_c)]$, resulting in a local group velocity[1] $v_g(\pmb{r}) = \Delta r/\Delta t = |\pmb{\nabla}[\partial_\omega \varphi(\pmb{r},\omega_c)]|^{-1}$. Observe that this group velocity, while being a function of the position variable $\pmb{r}$, has no time dependence.

**Example**. A Gaussian beam, whose waist of radius $R_0$ is located in the $xy$-plane at $z = 0$, has a real-valued spectral distribution $A_0(\omega)$ centered at $\omega = \omega_c$ with a narrow bandwidth of $\Delta\omega$. The initial amplitude distribution in the $z = 0$ plane is thus given by[6,16]

$$a_0(x,y,t) = \text{Re} \int_{\omega=0}^{\infty} A_0(\omega) \exp[-(x^2 + y^2)/R_0^2] e^{-i\omega t}d\omega. \tag{41}$$

The beam propagates along the positive z-axis within a homogeneous, isotropic, and transparent medium of refractive index $n(\omega)$. Considering that the Fourier transform of the initial Gaussian profile is $\pi R_0^2 \exp[-R_0^2(k_x^2 + k_y^2)/4]$, and that, in the paraxial regime, $k_z$ of Eq.(33) can be approximated as

$$k_z \cong (n\omega/c) - \tfrac{1}{2}(n\omega/c)^{-1}(k_x^2 + k_y^2), \qquad (42)$$

we obtain from Eq.(31) the following scalar amplitude profile throughout the entire spacetime:[16]

$$a(\mathbf{r},t) = \operatorname{Re} \int_{\omega=0}^{\infty} A_0(\omega)[1 + \mathrm{i}(2cz/n\omega R_0^2)]^{-1} e^{\mathrm{i}(n\omega/c)z} e^{-(x^2+y^2)/[R_0^2+\mathrm{i}(2cz/n\omega)]} e^{-\mathrm{i}\omega t} d\omega. \qquad (43)$$

A comparison with Eq.(35) now yields

$$\varphi(\mathbf{r},\omega) = \frac{n\omega z}{c} + \frac{(2cz/n\omega)(x^2+y^2)}{(2cz/n\omega)^2 + R_0^4} - \tan^{-1}\left(\frac{2cz}{n\omega R_0^2}\right). \qquad (44)$$

As a first step toward evaluating the local group velocity of the Gaussian wave-packet, we compute the gradient of $\varphi(\mathbf{r},\omega)$, as follows:

$$\boldsymbol{\nabla}\varphi(\mathbf{r},\omega) = \left(\frac{n\omega}{c}\right)\left\{\hat{\mathbf{z}} + \frac{(x/z)\hat{\mathbf{x}} + (y/z)\hat{\mathbf{y}} - \tfrac{1}{2}(R_0/z)^2\hat{\mathbf{z}}}{1+(n\omega R_0^2/2cz)^2} - \frac{1-(n\omega R_0^2/2cz)^2}{[1+(n\omega R_0^2/2cz)^2]^2}\left(\frac{x^2+y^2}{2z^2}\right)\hat{\mathbf{z}}\right\}. \qquad (45)$$

The local group velocity $\mathbf{v}_g(\mathbf{r})$ is obtained from the derivative with respect to $\omega$ of the above gradient, evaluated at the center frequency $\omega = \omega_c$. The calculation is straightforward but somewhat tedious. In what follows, we shall discuss only a few special cases.

i) In the far field, where $z \gg \pi n(\omega_c)R_0^2/\lambda_c$ (with $\lambda_c = 2\pi c/\omega_c$ being the vacuum wavelength at the center frequency), we find

$$\boldsymbol{\nabla}[\partial_\omega \varphi(\mathbf{r},\omega_c)] \cong \left[\left(\frac{x}{z}\right)\hat{\mathbf{x}} + \left(\frac{y}{z}\right)\hat{\mathbf{y}} + \left(1 - \frac{x^2+y^2+R_0^2}{2z^2}\right)\hat{\mathbf{z}}\right]\partial_\omega\left(\frac{n\omega}{c}\right)\bigg|_{\omega_c}. \qquad (46)$$

The group velocity in the far field is seen to be aligned with the direction of the local position vector $\mathbf{r}$, with a magnitude that approaches $c/\partial_\omega(n\omega)|_{\omega_c}$.

ii) At the waist of the beam, where $z = 0$, we find

$$\boldsymbol{\nabla}[\partial_\omega \varphi(\mathbf{r},\omega_c)] = \{1 + 2(c/n\omega_c R_0)^2[1 - (x^2+y^2)/R_0^2]\}\partial_\omega(n\omega/c)|_{\omega_c}\hat{\mathbf{z}}. \qquad (47)$$

The on-axis group velocity is seen to be reduced by a factor of $1 + 2(c/n\omega_c R_0)^2$ relative to that at the far field, namely, $c/\partial_\omega(n\omega)|_{\omega_c}$. Away from the center, however, the velocity rises with distance from the center, eventually exceeding the far field velocity by a small amount.

iii) Along the optical axis, where $x = y = 0$, we find

$$\boldsymbol{\nabla}[\partial_\omega \varphi(\mathbf{r},\omega_c)]|_{(x,y)=(0,0)} = \left[1 + \frac{(n\omega_c R_0/2c)^2 - (z/R_0)^2}{2[(n\omega_c R_0/2c)^2 + (z/R_0)^2]^2}\right]\partial_\omega\left(\frac{n\omega}{c}\right)\bigg|_{\omega_c}\hat{\mathbf{z}}. \qquad (48)$$

The on-axis group velocity peaks at $z = \pm\sqrt{3}\,\pi n(\omega_c)R_0^2/\lambda_c$, which is $\sqrt{3}$ times the Rayleigh range of the beam.[16] The peak value of $v_g$ at this point equals the far field velocity $c/\partial_\omega(n\omega)|_{\omega_c}$ divided by $1 - (c/2n\omega_c R_0)^2$.

**7. Concluding remarks**. We have examined the physical assumptions and mathematical methods and arguments that underlie some of the standard treatments of EM wave-packets propagating within LHI media. When a packet has a narrow spectral bandwidth as well as a $k$-vector content of limited angular spread, we found, under certain additional constraints, that one could arrive at simple formulas to trace the trajectory of the peak of the packet's envelope, and to estimate its associated group velocity. This was done for wave-packets residing in unbounded media as well as those propagating within waveguides or confined to certain surfaces and interfaces. Looking at these narrowband wave-packets from a somewhat different perspective, we also showed the possibility of tracking the evolution of their envelopes by examining their local group velocities at different positions in space. A natural extension of this work would relate the higher-order derivatives of the spatio-temporal spectrum of a wave-packet to changes in its other important characteristics in consequence of propagation through an LHI medium. Other directions for future investigations include addressing the effects of birefringence, optical activity, and spatial dispersion in conjunction with various nonlinearities of material media—with an eye toward applications in pulse compression, frequency comb generation, optical and microwave spectroscopy, fiber lasers, fiber amplifiers, and optical communication systems.

**Dedication**. This paper is dedicated to the memory of Roland V. Shack, who occupied an office two doors down the hall from mine, and was my colleague and beloved friend for nearly 25 years. Not only did I have the honor of knowing him as a neighbor, but also we had many technical (and some non-technical) discussions over the years. Some of my fondest memories of Roland pertain to conversations that he and I had on issues related to electromagnetic dispersion in material media as well as in certain optical systems. This was a topic of great interest to Roland, and I dare to harbor the hope and entertain the idea that my musings in this paper would not have disappointed him.